\documentclass[aps,prd,showpacs,amssymb,twocolumn,nofootinbib,
superscriptaddress]{revtex4}

\usepackage{amssymb,amsmath}
\usepackage{amsfonts}

\def\be{\begin{equation}}
        \def\ee{\end{equation}}
        \def\ba{\begin{eqnarray}}
        \def\ea{\end{eqnarray}}

\def\G{\Gamma}         
\def\R{\mathbb{R}}     
\def\N{\mathbb{N}}     
\def\Z{\mathbb{Z}}     

\begin{document}


\title{Unitary evolution in Gowdy cosmology}

\author{Alejandro Corichi}\email{corichi@matmor.unam.mx}
\affiliation{Instituto de Matem\'aticas, UNAM, A. Postal 61-3,
Morelia, Mich. 58090, Mexico} \affiliation{Instituto de Ciencias
Nucleares, UNAM, A. Postal 70-543, Mexico D.F. 04510, Mexico.}
\author{Jer\'onimo Cortez}\email{jacq@iem.cfmac.csic.es}
\affiliation{Instituto de Estructura de la Materia, CSIC, Serrano
121, 28006 Madrid, Spain.}
\author{Guillermo A. Mena Marug\'an}\email{mena@iem.cfmac.csic.es}
\affiliation{Instituto de Estructura de la Materia, CSIC, Serrano
121, 28006 Madrid, Spain.}

\begin{abstract}
Recent results on the non-unitary character of quantum time
evolution in the family of Gowdy $T^{3}$ spacetimes bring the
question of whether one should renounce in cosmology to the most
sacred principle of unitary evolution. In this work we show that the
answer is in the negative. We put forward a full nonperturbative
canonical quantization of the polarized Gowdy $T^{3}$ model that
implements the dynamics while preserving unitarity. We discuss
possible implications of this result.
\end{abstract}

\pacs{04.60.Ds, 04.62.+v, 04.60.Kz, 98.80.Qc}

\maketitle

The issue of unitarity in quantum gravity has been of central
importance for the past 30 years, ever since the possibility of
black holes evaporating suggested that unitary evolution might be
violated. However, most of the attention on this issue has been
within the semiclassical and minisuperspace scenarios where the
gravitational degrees of freedom are at most finite in number. It is
then quite natural to investigate the issue of time evolution within
the full theory, or at least for models that still possess an
infinite number of degrees of freedom. Of particular relevance are
cosmological, spatially closed, models where no canonical notion of
an (asymptotic) unitary time evolution exists (as is the case for
asymptotically flat and anti-de Sitter boundary conditions). It is
with this in mind that we consider the simplest of all inhomogeneous
closed models, namely the Gowdy $T^3$ cosmology.

Since the mid seventies, the quantization of the Gowdy $T^{3}$ model
\cite{gowdy} has received a great deal of attention
\cite{varios,berger1}. The first preliminary attempts
\cite{varios,berger1,berger} to define a quantum theory and extract
physics from the model were followed by more detailed analysis
\cite{guillermo-gowdy,husain-smolin}. Considerable progress has
recently been achieved in defining a complete quantization of the
(sub-)model with linear polarization \cite{pierri}.

The quantization proposed in \cite{pierri} is based on the fact
that the polarized model can be treated as $2+1$ gravity coupled
to an axially symmetric, massless scalar field, defined in a
manifold whose topology is $T^2 \times \R^+$. More precisely, once
the system is (partially) gauge fixed and a choice of internal
time is made, the spacetimes are characterized (modulo a remaining
global constraint) by a ``point particle" degree of freedom and a
free scalar field $\phi$ propagating in a fictitious
two-dimensional expanding torus. Thus, the problem of quantization
of the local gravitational degrees of freedom reduces to a quantum
theory of the scalar field in the fictitious background. The
quantum Gowdy $T^{3}$ model is defined by using a representation
for $\phi$ on a fiducial Fock space, where the remaining
constraint is imposed to get the physical Hilbert space.

Despite this progress, the quantization put forward in \cite{pierri}
has a serious drawback: the dynamics cannot be implemented as a
unitary transformation, neither on the kinematical \cite{ccq-t3} nor
in the physical \cite{torre-prd} Hilbert space. Even though the
dynamics can be approximated as much as desired by means of unitary
transformations \cite{jg}, the model is still lacking a unitary
operator that represents the genuine time evolution. The failure of
unitarity is, in the best of cases \cite{pe}, a non-trivial
complication that impedes the availability of a Schr\"odinger
picture in which dynamics preserves the conventional notion of
probability \cite{torre-prd,jacob}. The question then rises of
whether one should really abandon the concept of unitary evolution
or look instead for a different quantization of the model compatible
with unitarity. The relevance of this question surpasses the
restricted context of the Gowdy cosmology, which can be viewed as a
particular arena in which one is addressing the issue. The aim of
this work is to show that, opposite to any pessimistic perspective,
it {\it is} possible to achieve a unitary quantum dynamics in the
polarized Gowdy cosmology. At least as far as this system is
concerned, there is no intrinsic obstruction to the standard
probabilistic interpretation of quantum physics (neither in the
Heisenberg nor in the Schr\"odinger picture) in a cosmological
scenario.

Let us briefly recall the model introduced in \cite{pierri}, which
was essentially constructed starting with a (partially) gauge-fixed
system \cite{jg} that, modulo a global constraint, consists of a
reduced phase space ${\G}_{r}=\G_{0}\oplus \tilde{\G}$, where
$\G_{0}$ and $\tilde{\G}$ admit as respective coordinates a ``point
particle" canonical pair $(Q,P)$ and a ``field'' canonical pair
$\big(\phi(t,\theta) , P_{\phi}(t,\theta)\big)$. Here, $\phi$ and
$P_{\phi}$ are functions of the (internal) time coordinate $t$ and
the spatial coordinate $\theta \in S^{1}$. The corresponding reduced
Hamiltonian is \be \label{reduced-ham} H_{r}=\frac{1}{2} \oint
d\theta \, \left(\frac{P_{\phi}^{2}}{t}+t\phi'\, ^{2}\right). \ee
Thus, the ``point particle" degrees of freedom are constants of
motion, whereas a nontrivial evolution takes place only in the field
sector $\tilde{\G}$. To be more precise, $\phi$ must satisfy the
second-order differential equation \be \label{eq-phi}
\ddot{\phi}+\frac{\dot{\phi}}{t}-\phi''=0 . \ee All smooth solutions
to (\ref{eq-phi}), that we will generically denote by $\varphi$, can
be written as \be \label{scalarfield-sol}
\varphi(t,\theta)=\sum_{n\in
\Z}\left[A_{n}f_{n}(t,\theta)+A^{*}_{n}f^{*}_{n}(t,\theta)\right],\ee
with \be f_{0}(t,\theta)= \frac{1-i\ln t}{\sqrt{4\pi}},\quad
f_{n}(t,\theta)=\frac{H_{0}(|n|t)}{\sqrt{8}}e^{in\theta}\quad n\neq
0.\nonumber \ee The symbol $*$ denotes complex conjugation, $H_{0}$
is the zeroth-order Hankel function of the second kind
\cite{abramowitz}, and in order to guarantee pointwise convergence,
the sequence of constant coefficients $\{A_{n}\}$ must decrease
faster than the inverse of any polynomial in $n$ as $n\to \pm
\infty$.

Equation (\ref{eq-phi}) is the Klein-Gordon equation for a massless,
axially symmetric free scalar field propagating in a fictitious
background $\big({\cal{M}}\simeq T^{2}\times
\R^{+},g^{(f)}_{ab}\big)$, where \be
g^{(f)}_{ab}=-(dt)_{a}(dt)_{b}+(d\theta)_{a}(d\theta)_{b}
+t^{2}(d\sigma)_{a}(d\sigma)_{b},\nonumber \ee with $t\in \R^{+}$
and $\theta ,\sigma \in S^1$ \cite{note4}. Hence, we can identify
$\tilde{\G}$ with the canonical phase space of the field in this
background, while the space $\tilde{S}$ of smooth solutions can be
considered as the covariant phase space of this Klein-Gordon field.
Endowing the space $\tilde{S}$ ($\tilde{\Omega}$ being its
symplectic form) with the ``natural" $\tilde{\Omega}$-compatible
complex structure $\tilde{J}$ \cite{pierri}: \be \label{pierris-cs}
\tilde{J}(\bar{f}_{n})=i\bar{f}_{n}, \quad
\tilde{J}(\bar{f}^{*}_{n})=-i\bar{f}^{*}_{n} , \ee where
$\bar{f}_{n}(t):=f_{n}(t,\theta)\exp[-in\theta]$, one can construct
the ``one-particle" Hilbert space ${\tilde{\cal{H}}}$ and, from it,
the symmetric Fock space ${\cal{F}}(\tilde{\cal{H}})$ on which the
formal field operator is written in terms of creation and
annihilation operators [corresponding to the positive and negative
frequency decomposition defined by the complex structure
(\ref{pierris-cs})]. However, the Bogoliubov transformation that
implements the dynamics in the quantum theory --by relating at
different times either states in $\tilde{\cal{H}}$ \cite{ccq-t3}
(Schr\"{o}dinger picture) or creation and annihilation operators
\cite{torre-prd} (Heisenberg picture)-- turns out not to be square
summable in its antilinear part. As a consequence, the evolution
dictated by the Hamiltonian (\ref{reduced-ham}) fails to be
unitarily implementable at the kinematical level \cite{ccq-t3} as
well as in the physical Hilbert space \cite{torre-prd}. This ends
our brief review of the current status of the quantization proposed
in \cite{pierri}.

In order to arrive at a unitary theory, we will use the freedom
available to redefine the classical phase space through
time-dependent canonical transformations. With the resulting set of
new canonical variables and its corresponding Hamiltonian, one may
then reformulate the quantum Gowdy model. Thus, let us consider the
specific canonical transformation \cite{remark}: \be
\label{cano-transf} \bar{Q}:=Q, \quad \bar{P}:=P, \quad
\xi:=\sqrt{t}\phi, \quad
P_{\xi}:=\frac{P_{\phi}}{\sqrt{t}}+\frac{\phi}{2\sqrt{t}}. \ee
Taking into due account the explicit time-dependence of this
transformation, the reduced Hamiltonian for the new system of
variables becomes \be \label{hamil-new-var} \bar{H}_{r}=
\frac{1}{2}\oint d\theta
\left(P_{\xi}^{2}+\xi'^{2}+\frac{\xi2}{4t^{2}} \right). \ee Note
that this is the Hamiltonian of an axially symmetric, free scalar
field with a time-dependent potential that represents an effective
mass $1/(2t)$, propagating in a fictitious static background
$({\cal{M}}\approx T^{2}\times \R^{+} , \bar{g}^{(f)}_{ab})$ with
\be \bar{g}^{(f)}_{ab}=-(dt)_{a}(dt)_{b}+
(d\theta)_{a}(d\theta)_{b}+(d\sigma)_{a}(d\sigma)_{b}.\nonumber \ee

The Hamiltonian equations derived from (\ref{hamil-new-var}) lead to
\be \label{scalarf.new-var} \ddot{\xi}-\xi''+\frac{\xi}{4t^{2}}=0.
\ee We will denote by $\zeta$ the smooth solutions to
(\ref{scalarf.new-var}), which adopt the generic form \be
\label{scalarf-in-a-var} \zeta(t,\theta)=\sum_{n\in
\Z}\left[A_{n}g_{n}(t,\theta)+A_{n}^{*}g_{n}^{*}(t,\theta)\right],
\ee where $g_{n}(t,\theta):=\sqrt{t}f_{n}(t,\theta)$, as it is clear
from (\ref{scalarfield-sol}) and (\ref{cano-transf}). The complete
set of mode solutions $\{g_{n}(t,\theta)\}$ is ``orthonormal" in the
product $(g_{l},g_{n})=-i\Omega(g_{l}^{*},g_{n})$ [i.e.
$(g_{l},g^{*}_{n})=0$,
$(g_{l},g_{n})=\delta_{ln}=-(g^{*}_{l},g^{*}_{n})$], with
\cite{note2} \be \Omega(\zeta_{1},\zeta_{2})=\oint d\theta \,
\left(\zeta_{2}
\partial_{t}\zeta_{1}-\zeta_{1} \partial_{t}\zeta_{2}\right). \nonumber
\ee Hence, in the field sector, the covariant phase space is the
symplectic vector space $S:=(\Omega,\{\zeta\})$, which can equally
be coordinatized by the (pairs of complex conjugate) variables
$\{(A_{n},A^{*}_{n})_{n\in \Z}\}$.

Alternatively, we can consider the canonical phase space $\G$,
coordinatized by the set of (complex) canonically conjugate pairs
$\{(\xi_{n},P_{\xi}^{-n})_{n\in \Z}\}$, where $\xi_{n}$ and
$P_{\xi}^{n}$ are the (implicitly time-dependent) Fourier
coefficients of the configuration and momenta of the massive scalar
field. Let us now introduce the following transformations for the
zero and nonzero modes, respectively: \ba
b_{0}=\frac{\xi_{0}+iP_{\xi}^{0}}{\sqrt{2}} &,&\quad b_{0}^{*}
=\frac{\xi_{0}-iP_{\xi}^{0}}{\sqrt{2}}, \nonumber \\
b_{n}=\frac{|n| \xi_{n}+iP_{\xi}^{n}}{\sqrt{2|n|}} &,& \quad
b_{-n}^{*}=\frac{|n| \xi_{n}-iP_{\xi}^{n}}{\sqrt{2|n|}}. \nonumber
\ea They are canonical, inasmuch as
$\{b_{n},ib_{m}^{*}\}=\delta_{nm}$. So, one can adopt as coordinates
for $\G$ the (complex conjugate) variables
$\{(b_{n},b^{*}_{n})_{n\in \Z}\}$.

The map from $S$ to $\G$ is given by
\ba b_{0}(t) & =& r_{0}(t)A_{0}+s_{0}(t)A^{*}_{0}, \nonumber \\
b_{n}(t) & = & c(x_{n})A_{n}+d(x_{n})A^{*}_{-n}, \label{map-a-to-b}
\ea (for the zero and nonzero modes) where \ba
s_{0}(t)&=&\sqrt{\pi}g^{*}_{0}(t,\theta)
\left(1+\frac{i}{2t}\right)-\frac{1}{2\sqrt{t}} , \nonumber \\
r_{0}(t)&=& 2\sqrt{\pi}g_{0}(t,\theta)-s^{*}_{0}(t) , \nonumber \ea
and \ba d(x_{n})&=& \sqrt{\frac{\pi x_{n}}{8}} \left[
\left(1+\frac{i}{2x_{n}}\right)H^{*}_{0}(x_{n})-iH^{*}_{1}(x_{n})
\right], \nonumber \\
c(x_{n})&=&\sqrt{\frac{\pi x_{n}}{2}} H_{0}(x_{n})- d^{*}(x_{n}).
\nonumber\ea Here, $x_{n}:=|n|t$ and $H_{1}$ is the first-order
Hankel function of the second kind \cite{abramowitz}. It is not
difficult to see that \be |r_{0}(t)|^{2}-|s_{0}(t)|^{2}=1,\quad
\nonumber |c(x_{n})|^{2}-|d(x_{n})|^{2}=1 , \nonumber\ee  for all
$t>0$. This reflects the fact that (\ref{map-a-to-b}) is a
Bogoliubov transformation between annihilation and creation-like
variables. Moreover, it can be shown that this time-dependent
canonical transformation is generated by the Hamiltonian
(\ref{hamil-new-var}). From (\ref{map-a-to-b}), it follows that in
the canonical phase space a state $(b_{n}(t_{0}),b^{*}_{n}(t_{0}))$
at time $t_{0}$ evolves to the state $(b_{n}(t),b^{*}_{n}(t))$ at
time $t$ according to \be \label{b-evol}
b_{n}(t)=\alpha_{n}(t,t_{0})b_{n}(t_{0})+
\beta_{n}(t,t_{0})b^{*}_{-n}(t_{0}), \ee where, for the nonzero
modes,
 \ba \alpha_{n}(t,t_{0}) & = &
c(x_{n})c^{*}(x^{0}_{n})-
d(x_{n})d^{*}(x^{0}_{n}), \nonumber \\
\beta_{n}(t,t_{0}) & = & d(x_{n})c(x^{0}_{n})-c(x_{n})d(x^{0}_{n}) ,
\nonumber\ea with $x_{n}^0:=|n|t_{0}$. For the zero modes, one gets
a similar expression, with the functions $c$ and $d$ substituted by
$r_0$ and $s_0$, and the arguments $x_n$ and $x_n^0$ replaced with
$t$ and $t_0$. We note that
$|\alpha_{n}(t,t_{0})|^{2}-|\beta_{n}(t,t_{0})|^{2}=1$ for all $n$,
as it should be because the map (\ref{b-evol}) on $\G$ is given by
the composition of two Bogoliubov transformations and therefore is
itself a transformation of this kind.

Taking a fixed time $t_{0}>0$ and considering the inverse of
(\ref{map-a-to-b}), expression (\ref{scalarf-in-a-var}) can be
written in terms of a new ``orthonormal" set of solutions
$\{G_{n}(t,\theta)\}$: \be \label{scalarf-ref-to}
\zeta(t,\theta)=\sum_{n \in
\Z}\left[G_{n}(t,\theta)b_{n}(t_{0})+G_{n}^{*}(t,\theta)b_{n}^{*}(t_{0})
\right].\ee  Explicitly, for the zero and nonzero modes, \ba
G_{0}(t,\theta)&=&\sqrt{t}\left[r_{0}^{*}(t_{0})f_{0}(t,\theta)-
s_{0}^{*}(t_{0})f^{*}_{0}(t,\theta)\right],\nonumber \\
G_{n}(t,\theta)&=& \sqrt{\frac{t}{8}}\left[c^{*}(x_{n}^{0})
H_0(x_{n})-d^{*}(x_{n}^{0}) H^*_0(x_{n})\right]e^{i n \theta}.
\nonumber \ea

Using the fact that the solutions in (\ref{scalarf-ref-to}) are
decomposed in complex conjugate pairs, we define a
$\Omega$-compatible complex structure $J$ as \be
J(\bar{G}_{n}(t))=i\bar{G}_{n}(t),\quad
J(\bar{G}^{*}_{n}(t))=-i\bar{G}^{*}_{n}(t),\nonumber\ee where
$\bar{G}_{n}(t):=G_{n}(t,\theta)\exp[-i n \theta]$. With
$(\Omega,\{\zeta\},J)$ we can construct the ``one-particle" Hilbert
space ${\cal{H}}$ and the associated symmetric Fock space
${\cal{F}}({\cal{H}})$, which will be the (kinematical) Hilbert
space of the quantum theory. Following this prescription, we can
introduce the formal field operator $\hat{\zeta}$ in terms of
creation and annihilation operators corresponding to the positive
and negative frequency decomposition provided by the complex
structure $J$: \be \hat{\zeta}(t;\theta)=\sum_{n \in
\Z}\left[G_{n}(t,\theta)\hat{b}_{n}+G_{n}^{*}(t,\theta)
\hat{b}_{n}^{\dagger}\right]. \nonumber\ee A comparison with
(\ref{scalarf-ref-to}) shows that we could have obtained this field
operator by naively promoting the constants of motion
$\{b_{n}(t_{0}),b_{n}^{*}(t_{0})\}$ in the solution to annihilation
and creation operators
$\{\hat{b}_{n}(t_{0})=\hat{b}_{n},\hat{b}_{n}^{\dagger}(t_{0})=
\hat{b}^{\dagger}_{n}\}$. This can be understood as the
Schr\"{o}dinger picture.

In the Heisenberg picture, time evolution is provided by the
Bogoliubov transformation (\ref{b-evol}). Namely, by calling
$\hat{b}^{(H)}_{n}(t_{0}):=\hat{b}_{n}$, the relation between the
annihilation and creation operators at different times $t_0$ and $t$
is \be  \hat{b}^{(H)}_{n}(t) =
\alpha_{n}(t,t_{0})\hat{b}^{(H)}_{n}(t_{0})+
\beta_{n}(t,t_{0})\hat{b}_{-n}^{(H)\dagger}(t_{0}) . \nonumber\ee
Thus, in this picture we get \be
\hat{\zeta}(t;\theta)=\frac{1}{\sqrt{4\pi}}\sum_{n\in
\Z}N_{n}\left[e^{in \theta}\hat{b}^{(H)}_{n}(t)+e^{-i n
\theta}\hat{b}_{n}^{(H)\dagger}(t)\right], \nonumber\ee where
$N_{n}=1/\sqrt{|n|}$, except for $N_{0}=1$.

Time evolution is unitarily implementable on the (kinematical) Fock
space ${\cal{F}}({\cal{H}})$ if and only if the sequence
$\{\beta_{n}\}$ is square summable \cite{shale}. Since
$\beta_{n}=\beta_{-n}$, it suffices to analyze the sequence
$\{\beta_{k}\}$ with $k\in \N$. From the large-argument asymptotic
expansions of the Hankel functions \cite{abramowitz} one can check
that, given any fixed $T>0$, the sequence $\{d(kT)\}$ (with $k\in
\N-\{0\}$) is square summable. In particular, so are $\{d(kt)\}$ and
$\{d(kt_{0})\}$. This implies that there exists an integer $k_{0}$
such that both $|d(kt)|$ and $|d(kt_{0})|$ are smaller than the
unity for all $k>k_{0}$. Since $|c|^{2}=1+|d|^{2}$, one also has
that $|c(kt)|$ and $|c(kt_{0})|$ are smaller than $\sqrt{2}$ for
$k>k_{0}$. In this case, \ba |\beta_{k}(t,t_{0})|^{2} &\leq& 2
\left(|d\big(x^{0}_{k}\big)|+|d(x_{k})|\right)^{2} \nonumber\\
&\leq& 4 \left(|d\big(x^{0}_{k}\big)|^{2}+|d(x_{k})|^{2}\right).
\nonumber\ea The square summability of the sequence $\{\beta_{k}\}$
follows then from that of $\{d(kt)\}$ and $\{d(kt_{0})\}$.

Time evolution is hence unitarily implementable on the (kinematical)
Fock space ${\cal{F}}({\cal{H}})$. Moreover, the evolution leaves
invariant the constraint \be \hat{\cal{C}}=\sum_{k\in
\N}k\left(\hat{b}^{\dagger}_{k}\hat{b}_{k}-
\hat{b}^{\dagger}_{-k}\hat{b}_{-k}\right),\nonumber \ee which
implements quantum mechanically the requirement that the total
($\theta$-)momentum of the field $\xi$ vanish \cite{jg}. As a
consequence, the dynamics is unitarily implementable not only on
${\cal{F}}({\cal{H}})$, but also on the physical Hilbert space
${\cal{F}}_{\rm{phys}}({\cal{H}})$, defined as the kernel of the
above constraint. Thus, we have achieved a quantization of the Gowdy
$T^3$ model where physical states (as well as operators) evolve in a
unitary way. This is our main result.

In what follows, we discuss some consequences of this quantization
and compare it with previous ones. The first remark is that, whereas
$\hat{\xi}(t_{0};\theta)$ evolves unitarily to
$\hat{\xi}(t;\theta)$, the formal operator
$\hat{\phi}(t;\theta):=\hat{\xi}(t;\theta)/\sqrt{t}$ (considered in
\cite{pierri}) does not. Namely, $\hat{\phi}(t_0;\theta)$ and
$\hat{\phi}(t;\theta)$ are not unitarily related. Thus, a suitable
choice of the fundamental field seems very important for a
consistent quantization. In making this choice, we have employed the
freedom available to redistribute the time dependence [via the
time-dependent transformation (\ref{cano-transf})] in an implicit
part, generated by the reduced Hamiltonian of the model, and an
explicit part (the factor $1/\sqrt{t}$ in $\phi$), whose time
variation does not need to be described by means of a unitary
transformation. Note also that it is natural to consider
time-dependent canonical transformations in the Gowdy model, since
the reduced Hamiltonian obtained by gauge fixing is already
explicitly time-dependent.

Let us now explore the new description proposed here from the
viewpoint of the classical structure needed to find the quantum
representation. It is known \cite{torre-vara} that time evolution
between two Cauchy surfaces, say $t=t_{1}$ and $t=t_{2}$, can be
described on the space of solutions $S$ by a uniparametric family of
symplectic maps ${\cal{T}}_{t_{1},T}$, with $T\in [t_{1},t_{2}]$. We
can consequently obtain a uniparametric family of induced complex
structures $J_{T}={\cal{T}}_{t_{1},T}J{\cal{T}}_{t_{1},T}^{-1}$. In
fact, since the dynamics is unitarily implementable, $J_{T}-J$ is
Hilbert-Schmidt (HS) on ${\cal{H}}$ for every element of this
family. Similarly, for the description discussed in \cite{pierri} in
terms of the field $\phi$, one gets a family
$\tilde{J}_{T}=\tilde{\cal{T}}_{t_{1},T}\tilde{J}
\tilde{\cal{T}}_{t_{1},T}^{-1}$ on $\tilde{S}$. However,
$\tilde{J}_{T}-\tilde{J}$ fails to be HS on $\tilde{\cal{H}}$ for
$T\neq t_{1}$. In addition, recall that the correspondence
$\varphi=\zeta/\sqrt{t}$ defines a map $R:S\to \tilde{S}$ between
both spaces of solutions. We can then consider the complex
structures $J_{T}'=RJ_{T}R^{-1}$ induced on $\tilde{S}$ by this map,
and ask whether $J_{T}'-\tilde{J}_{T}$ is HS on $\tilde{\cal{H}}$.
It is easy to show that this happens only for $T=t_{1}$ (i.e, for
$RJR^{-1}$ and $\tilde{J}$). Therefore, even when the complex
structures $J$ and $\tilde{J}$ are ($R$-)equivalent, they do not
connect unitarily equivalent theories in the explained sense,
because the evolution is unitary in one case (in the quantum theory
with fundamental field $\xi$) but not in the other (in the
quantization in which $\phi$ is viewed as the fundamental field). It
would also be interesting to explore different quantum
representations for each of these choices of fundamental field and
study their equivalence and the unitary implementability of the
dynamics \cite{cocome}.

Our results are worthy of some specific comments from the
perspective of quantum field theory in curved spacetimes. The Gowdy
model was described in \cite{pierri} in terms of a free massless
scalar field $\phi$ on a flat, but time-dependent background. A
conventional quantization of this field leads to a non-unitary
theory. By a field redefinition, which involves the time parameter,
we have mapped the system into a scalar field $\xi$ subject to a
time-dependent potential (interpreted as a time-dependent mass),
though now the background is flat and time independent (like
3-dimensional Minkowski spacetime, except for the topology). The
natural quantization of this new field that we have put forward
provides a theory in which the dynamics is unitary. Besides, one can
check that $J_{T}$ approaches, for asymptotically large values of
$T$, the Poincar\'e-invariant complex structure of Minkowski
spacetime in the limit in which the system becomes massless. Note
that it is only in that limit that the system is invariant under
time-translations. This result points towards a possible connection
between unitary implementability and asymptotic symmetries, but the
matter certainly calls for a more thorough investigation.

Finally, let us point out that the vacuum of the quantum theory
proposed for $\xi$ is not left invariant by the time evolution, as
can be seen either from the presence of the time-dependent potential
in the reduced Hamiltonian or from the fact that the induced complex
structures $J_{T}$ differ from $J$ except at $T=t_1$. The quantum
description provides in this way a Hilbert space of physical states
${\cal F}_{\rm phys}({\cal H})$, obtained from a kinematical Fock
space ${\cal F}({\cal H})$, in which the vacuum evolves approaching
the Poincar\'e-invariant vacuum associated with the asymptotic
region at large times. This evolution may be interpreted as a
production of ``particles'' by the vacuum \cite{cocome,note3}.
Nevertheless, no conflict arises for unitarity. In conclusion, we
have succeeded in constructing a consistent quantization of the
polarized Gowdy model in which, on the one hand, a notion of vacuum
that displays a cosmological evolution is available and, on the
other hand, dynamics is unitary. This is the first cosmological
model with an infinite number of degrees of freedom for which a
quantization with these features has been constructed.

This work was supported by the Spanish MEC projects FIS2004-01912
and HP2003-0140, by CONACyT (M\'exico)
 40035-F and U47857-F, and by DGAPA-UNAM IN108103 grants. J.C. was
funded by the Spanish MEC, No./Ref. SB2003-0168.


\begin{thebibliography}{99}
\bibitem{note4}Note that the field is invariant under a $U(1)$ spatial
symmetry, since it depends on the angular coordinate $\theta$.
\bibitem{remark} The change $\xi=\sqrt{t}\phi$ was already
considered in \cite{berger1}, but just within the study of the WKB
regime.
\bibitem{note2} This symplectic form can be obtained, e.g., from the
action functional of the model once the canonical transformation
(\ref{cano-transf}) has been performed, and is independent of the
constant-$t$ section in which it is evaluated.
\bibitem{note3} It is worth pointing out that our quantization
scheme is radically different from any kind of instantaneous
diagonalization. The explicit time dependence of both the canonical
transformation (\ref{cano-transf}) and the reduced Hamiltonian
(\ref{hamil-new-var}) has been taken into full account.
\end{thebibliography}
\end{document}